\documentclass[12pt,a4paper]{article}

\usepackage{url,hyperref}
\usepackage{amsmath}
\usepackage{array,multirow,graphicx}
\usepackage{color}
\usepackage{anyfontsize}
\usepackage{textcomp}
\usepackage[left=2cm, right=2cm, top=2cm, bottom=2cm]{geometry}

\title{High-resolution computer meshes of the lower body bones of an adult human female derived from CT images}
\author{Manish Sreenivasa\thanks{manishs@uow.edu.au, University of Wollongong, Australia} and Daniel Gonzalez-Alvarado\thanks{Heidelberg University, Germany}}
\date{ver. 13.07.2018}

\begin{document}
\maketitle

\begin{abstract}
Computer-based geometrical meshes of bones are important for
applications in computational biomechanics as well as clinical software. There is however a lack of freely available detailed bone meshes, especially related to the human female morphology. 

We provide high resolution bone meshes of the lower body, derived from CT images of a 59 year old female cadaver that were sourced from the Visible Human Data Set, Visible Human Project (NIH, USA). Important bone landmarks and joint rotation axes are identified from the extracted meshes. A script-based framework is developed to provide a graphical user interface  that can visualize, resample and modify the meshes to fit different subject scales.

This open-data resource fills a gap in available data and is provided for free usage in research and other applications. The associated scripts allows users to easily transform the meshes to different laboratory and software setups. This resource may be accessed through the following web link: \url{https://github.com/manishsreenivasa/BMFToolkit}

{\bf  Please cite this work as: M Sreenivasa, D Gonzalez-Alvarado - arXiv preprint arXiv 1804.03655, 2018, DOI: 10.5281/zenodo.889060}

\end{abstract}

\section*{Introduction}
Computer-based bone meshes are useful for clinical illustrations as well as for applications in computational biomechanics such as patient-specific modeling \cite{Zhang2015,Sreenivasa2016}, and muscle simulations \cite{Delp2007}. Such meshes may be derived from post-processing of medical imaging data of live subjects or cadavers.

Datasets belonging to adult human males are freely available, for example as part of the Opensim software \cite{Delp2007,Arnold2010}, and the BodyParts3D dataset \cite{BodyParts3D}. Computer models of individual bones such as the pelvis (as well as the nearby organs and musculature) have also been made available for research and teaching purposes \cite{Sergovic2010,Kraima2013,Shin2013}. Commercial products such as those from Zygote (USA), provide detailed computer meshes of the human skeleton and other anatomical systems. To the best of our knowledge, there exists a lack of freely available high-resolution meshes of all bones of the lower body of an adult human female.
	
Here, we detail such an open-data resource derived from the Visible Human Project (VHP) dataset. We provide high resolution computer meshes of all lower body bones of an adult human female.  In addition we provide a software framework that allows users to visualize the meshes, manipulate and export them. The resulting meshes may be used as individual bones or as a linked kinematic model of the lower body skeleton.

\begin{figure}[h!]
  \centering
  \includegraphics[width=0.9\textwidth]{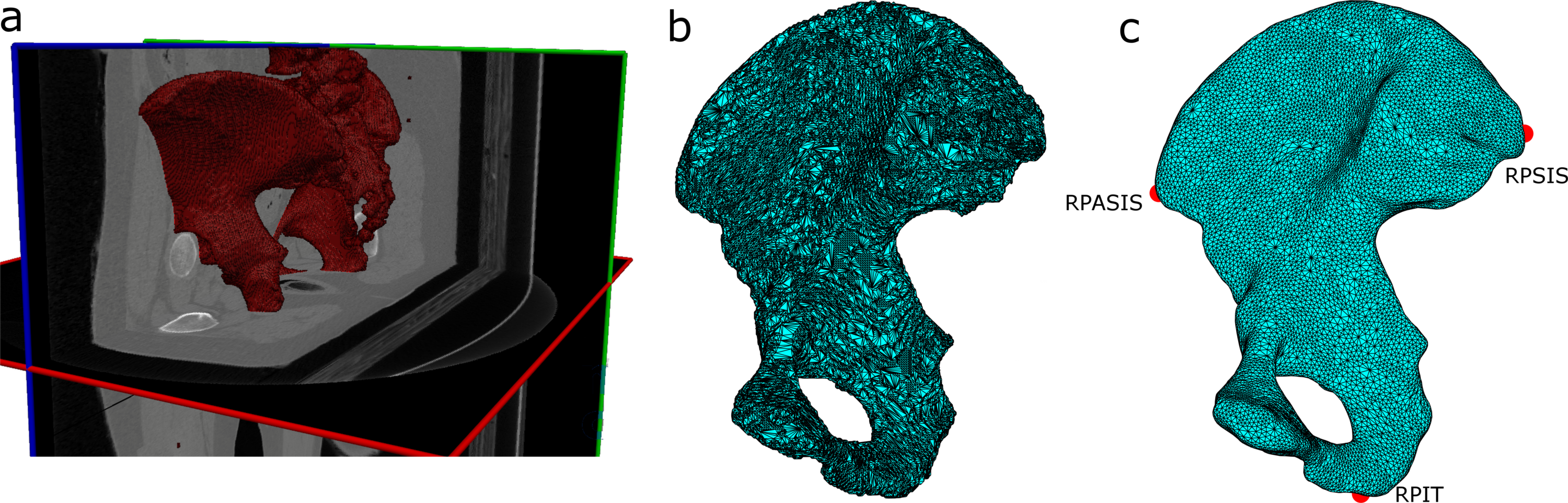}
  \caption{Processing of CT images.
      (a) Raw CT images were manually segmented in MITK \cite{Nolden2013}. (b) The resulting rough mesh was exported and (c) refined further to produce smooth bone meshes with landmark information.}
      	\label{fig:process}
\end{figure}

\section*{Methods}
In the following we describe the main steps to create the bone meshes from raw imaging data and the setup of the corresponding kinematic model.
\subsection*{Imaging data}
The Visible Human Project\textsuperscript{\tiny\textregistered}(VHP)\footnote{\url{https://www.nlm.nih.gov/research/visible/visible\_human.html}} provides datasets of male and female cadaveric images taken at high resolution \cite{VHPPaper,Spitzer1998}. For this work, we use the dataset corresponding to a 59 year old female. The methods used to extract the images are detailed elsewhere \cite{Spitzer1998}. The VHP dataset in DICOM format were sourced from the University of Iowa \footnote{\url{https://mri.radiology.uiowa.edu/visible_human_datasets.html}} with a pixel size of 0.33mm and inter-slice distance of 1.0mm.
\subsection*{Mesh Processing}
\subsubsection*{Image Segmentation}
\label{subsubsec:imageSeg}
Bone contours are segmented from DICOM images using the open-source software MITK 2014.10 (Fig. \ref{fig:process}a)\cite{Nolden2013}. Scans containing data for individual bones of the right lower body (see Table \ref{tab:landmarks} for list) are processed separately. The motivation behind only segmenting the right side bones is to have symmetric lower body meshes at the final step. We use the 3D Upper/Lower Threshold tool from the Segmentation Plugin in MITK to demarcate the bone contour in each scan. This initial estimate is then checked and corrected in each scan using 2D editing tools. The segmented contours are then used to develop 3D meshes using the Create Polygon Model tool, Fig. \ref{fig:process}a \& b. For the Sacrum, we define a sagittal clipping plane that splits the bone mesh into two halves, and use the right half for further processing. 
\subsubsection*{Mesh Refinement}
\label{subsubsec:meshRefine}
The rough 3D meshes are further processed in Matlab 2015b (The Mathworks Inc., USA) using the open-source toolbox Iso2Mesh \cite{Fang2009}. The following steps are consecutively applied to each of the bone meshes:
\begin{enumerate}
\item
Mesh Repair - Remove duplicated, isolated and self-intersecting elements
\item
Mesh Resample - Down-sample the repaired mesh
\item
Mesh Smooth - Low pass filter the down-sampled mesh
\end{enumerate}
For the Femur, Tibia and Fibula we further down-sample the mesh points pertaining to the bone shafts. This is done to reduce the mesh size without losing shape details as the bone shafts contain relatively sparse features. 

\begin{figure}[h!]
  \centering
  \includegraphics[width=0.9\textwidth]{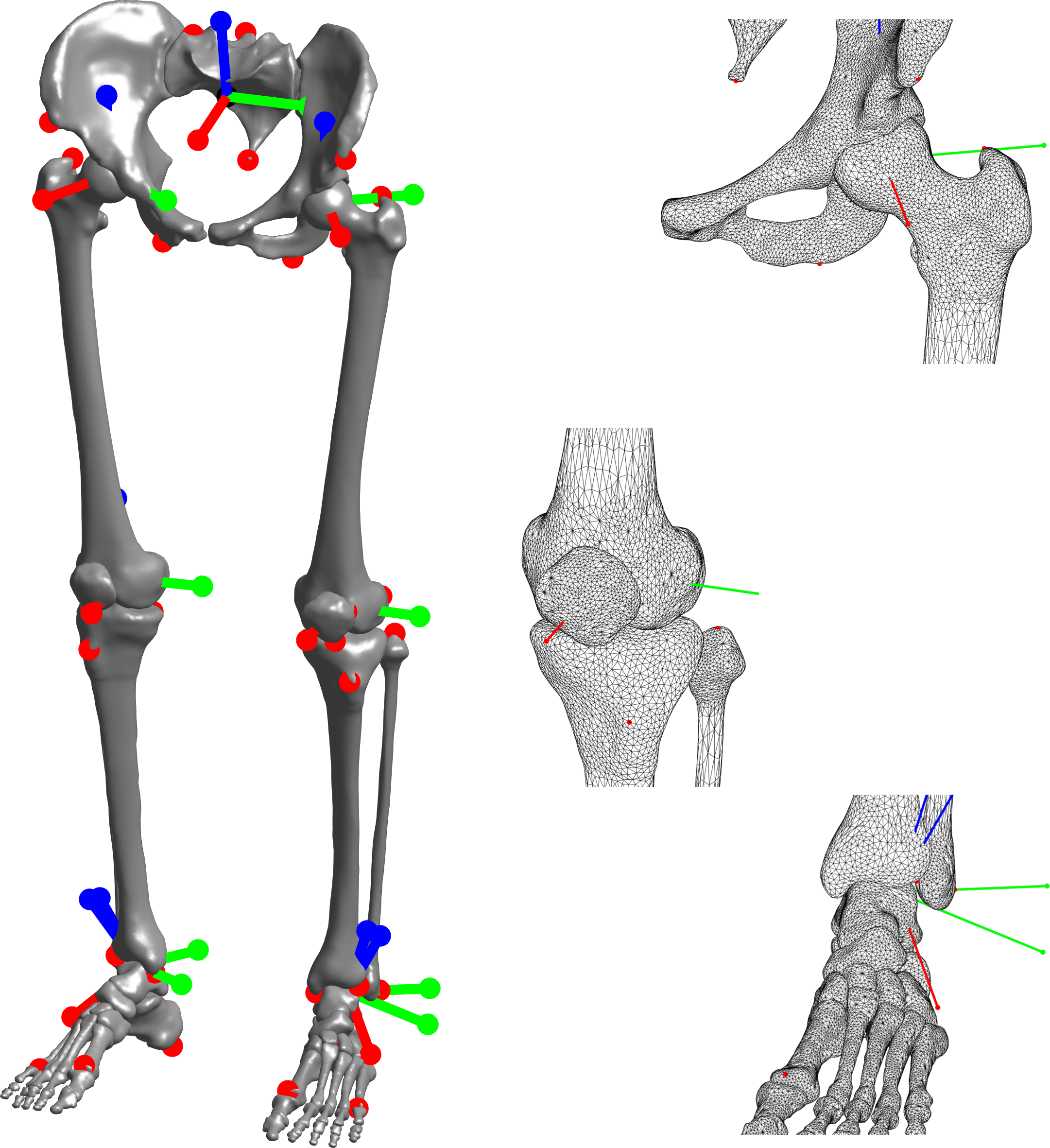}
  \caption{Processed meshes of the lower body skeleton.
      Insets show the detail levels at the hip, knee and ankle joints. Red circles indicate the identified landmarks. Joint axes are shown as orthogonal red (x-axis), green (y-axis) and blue (z-axis) lines.}
      \label{fig:allMeshes}
\end{figure}

Fig. \ref{fig:process}b-c shows a representative example of the final processed mesh for the right pelvic bone. The processed meshes are rotated to a neutral pose of the lower body skeleton, as shown in Fig. \ref{fig:allMeshes}. Adjacent mesh pairs are checked and corrected for bone-to-bone impingement. Finally, the meshes are mirrored about the previously identified sagittal plane to produce the full lower body skeletal bones. Mesh vertices corresponding to bony points of interest are identified as listed in Table \ref{tab:landmarks}.
\paragraph*{Known issues in original data}
\begin{itemize}
\item
A misalignment in the transverse plane center of the CT scans was corrected at the femoral mid-shaft level.
\item
One scan each at the femoral and tibial/fibular shaft levels were corrupted, and excluded from further processing.
\item
The neutral pose of the bones were visually approximated based on anatomical literature \cite{GrantsAtlas}. The bones of the feet required especially large corrections, possibly due to contractures. 
\item
An unusual bony growth, possibly an osteophyte, was detected on the distal end of the 1st metatarsal bone. This was corrected manually in the segmentation phase.
\end{itemize}

\subsection*{Kinematic Model}
Based on the identified bony landmarks we compute the static joint centers and rotation axes of the lower body skeleton. Femoral and ankle joint axes are defined as per Wu et. al \cite{Wu2002}, and the knee as per Grood \& Suntay \cite{Grood1983}. Fig. \ref{fig:allMeshes} plots all the lower body meshes, joint centers and rotation axes. 

The default kinematic model consists of a total of 12 internal degrees of freedom (DoF), with an additional 6 DoF on the sacrum as the free-flying root joint. The pelvis is the child of the sacrum and rigidly attached to it. The femur is the child of the pelvis and modeled as a 3-DoF spherical joint incorporating the extension-flexion, adduction-abduction and internal-external rotation of the hip. The patella is the child of the femur and is rigidly attached to it. The tibia is the child of the femur and modeled as a 1-DoF rotational joint incorporating flexion-extension of the knee. The fibula is the child of the tibia and is rigidly attached to it. The talus is the child of the tibia and modeled as a 1-DoF rotational joint incorporating flexion-extension of the ankle. The calcaneus is the child of the talus and modeled as a 1-DoF rotational joint incorporating inversion-eversion of the ankle. The remaining bones of the feet are rigidly attached to the calcaneus.

In addition to the bony landmarks we add virtual markers to the bone segments (see Table \ref{tab:landmarks}) as per an extended VICON Plugin-Gait markerset \cite{ViconPIG}. These markers may be used for the computation of generalized coordinates from recorded motion capture data and inverse kinematics analysis. Sample inverse kinematics results computed using the open-source software Puppeteer\footnote{\url{https://github.com/martinfelis/puppeteer}}\cite{Felis2015}, are included in the dataset. Note that the model and mesh dimensions defined so far corresponds to the  proportions of the VHP female subject. This \emph{original} model is provided in Lua\footnote{\url{https://www.lua.org/}} format (see supplementary files). The currently used kinematic definition may be easily extended (e.g. to incorporate additional translation between the femur and tibia) as needed for specific applications.
\section*{Software Implementation}
\begin{figure}[h!]
  \centering
  \includegraphics[width=0.9\textwidth]{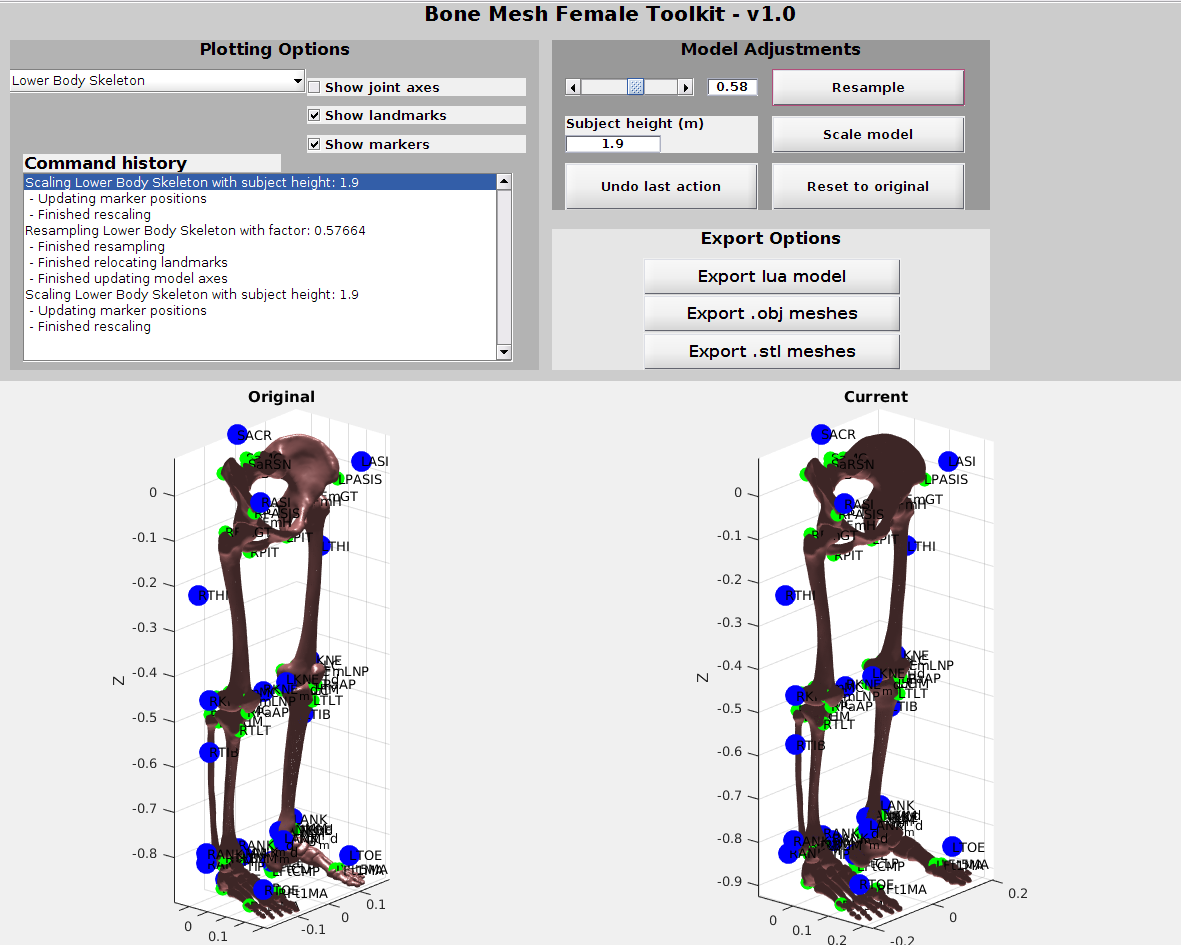}
  \caption{Mesh post-processing.
      Snapshot of the user interface with functionalities to resample, rescale and export meshes and associated model. Illustrated are the original model (left) and the result of the current pipeline (right). Green circles indicate the bony landmarks and blue circles the virtual markers.}
      \label{fig:GUI}
\end{figure}

The original meshes extracted from the raw data are incorporated within post-processing scripts \cite{BMFToolkit_v1}, written in Matlab/Octave (Fig. \ref{fig:GUI}). Users are provided a dropdown menu to choose to post-process all the lower-body bone meshes, or choose those of the feet, or any individual bone. The purpose of these scripts is to provide a simple user interface to scale the meshes to different body proportions based on anthropometric regression equations. We utilize the scaling provided by De Leva \cite{DeLeva1996} for this purpose. The user is asked to provide the overall height of the subject, and the meshes are linearly scaled based on the values from De Leva.

In addition, functionality is provided to resample the meshes at a lower resolution than that of the original set. The desired resolution is set relative to the original by using a slider or scalar value (Fig. \ref{fig:GUI}). This resampling functionality is provided to allow users to adapt the mesh detail to the specific application. For example, for studies that focus on a single bone a high level of detail may be necessary (e.g. femoral derotation), whereas for those regarding multi-joint movements (e.g. locomotion) a lower mesh detail is sufficient and faster to visualize. Mesh resampling is implemented by using the corresponding functions from the Iso2Mesh toolbox. 

The postprocessed meshes are visualized at each step to provide immediate feedback to the user. In addition, the meshes can be exported in OBJ and STL format, and the scaled kinematic model in Lua format. The model may be further used in animations using the open-source softwares Puppeteer \cite{Felis2015} and Meshup\footnote{\url{https://bitbucket.org/MartinFelis/meshup}}. A sample animation illustrating the degrees of freedom and range of motion of the model is available in the supplementary files to this article.

\section*{Conclusions}
We have presented detailed lower body bone meshes of an adult human female derived from the VHP dataset \cite{VHPPaper,Spitzer1998}. In addition, we provide a simple and fast user interface to scale and resample the meshes, and as well to create a kinematic model of the lower body. All data and code arising from this work are publicly accessible for further use in research and other applications \cite{BMFToolkit_v1}.

In future work we aim to provide further physiological detail by extracting the upper body bones, and as well, extend the functionality of the associated softwares. An important detail that may be useful would be to map the musculotendon insertion points on the bone meshes. Additionally, we can improve the current linear scaling of the meshes by incorporating subject-specific imaging data and using toolkits such as MRI2PSM \cite{Sreenivasa2016} to morph the bone meshes.

\subsection*{Availability of data and material}
The data generated during this study are included in this article's supplementary information files. Further developments are available in the BMFToolkit repository, \url{https://github.com/manishsreenivasa/BMFToolkit} \cite{BMFToolkit_v1}.

\subsection*{Funding}
This study was part of the Frontier-Orthosis project supported by the German Excellence Initiative within the third pillar funding of Ruprecht-Karls-Universit\"{a}t Heidelberg.

\subsection*{Authors' contributions}
MS developed the software architecture, and wrote the manuscript with assistance from DGA. DGA processed the raw data. Both authors read and approved the final manuscript.

\bibliographystyle{ieeetr}
\bibliography{bonemeshBib}

\begin{thebibliography}{10}

\bibitem{Zhang2015}
J.~Zhang, H.~Sorby, and T.~Besier, ``Musculoskeletal atlas project: Statistical
  model-based model generation,'' in {\em PMHA 2015 - 3rd International
  Workshop on Biomechanical and Parametric Modeling of Human Anatomy}, 2015.

\bibitem{Sreenivasa2016}
M.~Sreenivasa, C.~J.~G. Chamorro, D.~Gonzalez-Alvarado, O.~Rettig, and S.~I.
  Wolf, ``Patient-specific bone geometry and segment inertia from {MRI} images
  for model-based analysis of pathological gait,'' {\em Journal of
  Biomechanics}, vol.~49, no.~9, pp.~1918 -- 1925, 2016.

\bibitem{Delp2007}
S.~L. Delp, F.~Anderson, A.~Arnold, P.~Loan, A.~Habib, C.~John, E.~Guendelman,
  and D.~G. Thelen, ``Opensim: Open-source software to create and analyze
  dynamic simulations of movement,'' {\em IEEE Transactions in Biomedical
  Engineering}, vol.~54, pp.~1940--1950, 2007.

\bibitem{Arnold2010}
E.~M. Arnold, S.~R. Ward, R.~L. Lieber, and S.~L. Delp, ``A model of the lower
  limb for analysis of human movement,'' {\em Annals Biomedical Engineering},
  vol.~38, no.~2, pp.~269--279, 2010.

\bibitem{BodyParts3D}
N.~Mitsuhashi, K.~Fujieda, T.~Tamura, S.~Kawamoto, T.~Takagi, and K.~Okubo,
  ``Bodyparts3d: 3d structure database for anatomical concepts,'' {\em Nucleic
  Acids Research}, vol.~37, pp.~D782--5, 2009.

\bibitem{Sergovic2010}
A.~Sergovich, M.~Johnson, and T.~D. Wilson, ``Explorable three-dimensional
  digital model of the female pelvis, pelvic contents, and perineum for
  anatomical education,'' {\em Anatomical Sciences Education}, vol.~3,
  pp.~127--133, 2010.

\bibitem{Kraima2013}
A.~C. Kraima, N.~N. Smit, D.~Jansma, C.~Wallner, R.~L. A.~W. Bleys, C.~J. H.
  V.~D. Velde, C.~P. Botha, and M.~C. Deruiter, ``Toward a highly-detailed 3d
  pelvic model: Approaching an ultra-specific level for surgical simulation and
  anatomical education,'' {\em Clinical Anatomy}, vol.~26, pp.~333--383, 2013.

\bibitem{Shin2013}
D.~S. Shin, H.~G. Jang, S.~B. H. D.-H. Har, Y.~L. Moon, and M.~S. Chung,
  ``Two-dimensional sectioned images and three-dimensional surface models for
  learning the anatomy of the female pelvis,'' {\em Anatomical Sciences
  Education}, vol.~6, pp.~316--323, 2013.

\bibitem{Nolden2013}
M.~Nolden, S.~Zelzer, A.~Seitel, D.~Wald, M.~M{\"u}ller, A.~M. Franz,
  D.~Maleike, M.~Fangerau, M.~Baumhauer, L.~Maier-Hein, K.~H. Maier-Hein, H.~P.
  Meinzer, and I.~Wolf, ``The medical imaging interaction toolkit: challenges
  and advances,'' {\em International Journal of Computer Assisted Radiology and
  Surgery}, vol.~8, no.~4, pp.~607--620, 2013.

\bibitem{VHPPaper}
M.~J. Ackerman, ``The {Visible Human Project}
  \textsuperscript{\tiny\textregistered}: A resource for education,'' {\em
  Academic Medicine}, vol.~74, no.~6, pp.~667--670, 1999.

\bibitem{Spitzer1998}
V.~M. Spitzer and D.~G. Whitlock, ``The visible human dataset: The anatomical
  platform for human simulation,'' {\em The Anatomical Record (New Anat.)},
  vol.~253, pp.~49--57, 1998.

\bibitem{Fang2009}
Q.~Fang and D.~Boas, ``Tetrahedral mesh generation from volumetric binary and
  gray-scale images,'' in {\em IEEE International Symposium on Biomedical
  Imaging}, pp.~1142--1145, 2009.

\bibitem{GrantsAtlas}
A.~M.~R. Agur and A.~F. Dalley, {\em Grant's Atlas of Anatomy, 13th edition}.
\newblock Wolters Kluwer, Lippincott Williams \& Wilkins, 2013.

\bibitem{Wu2002}
G.~Wu, S.~Siegler, P.~Allard, C.~Kirtley, A.~Leardini, D.~Rosenbaum,
  M.~Whittle, D.~D. D’Lima, L.~Cristofolini, H.~Witte, O.~Schmid, and
  I.~Stokes, ``{ISB} recommendation on definitions of joint coordinate system
  of various joints for the reporting of human joint motion - {Part I}: ankle,
  hip, and spine,'' {\em Journal of Biomechanics}, vol.~35, no.~4,
  pp.~543--548, 2002.

\bibitem{Grood1983}
E.~S. Grood and W.~J. Suntay, ``A joint coordinate system for the clinical
  description of three-dimensional modtions: Application to the knee,'' {\em
  Journal of Biomechanical Engineering}, vol.~105, pp.~136--144, 1983.

\bibitem{ViconPIG}
Vicon, ``Lower body modeling with plug-in gait.''
  https://docs.vicon.com/display/Nexus25/Lower+body+modeling+with+Plug-in+Gait.

\bibitem{Felis2015}
M.~L. Felis, K.~Mombaur, and A.~Berthoz, ``An optimal control approach to
  reconstruct human gait dynamics from kinematic data,'' in {\em IEEE-RAS 15th
  International Conference on Humanoid Robots (Humanoids)}, pp.~1044--1051,
  2015.

\bibitem{BMFToolkit_v1}
M.~Sreenivasa, ``manishsreenivasa/bmftoolkit: Initial,'' Sept. 2017.

\bibitem{DeLeva1996}
P.~{de Leva}, ``Adjustments to {Zatsiorsky-Seluyanov's} segment inertia
  parameters,'' {\em Journal of Biomechanics}, vol.~29, no.~9, pp.~1223--1230,
  1996.

\end{thebibliography}

\begin{table}
\small
\caption{List of segmented bones, bony landmarks and markers. Only information for the right (R) side bones are listed.} \label{tab:landmarks} 
\begin{tabular}{|c|c|c|c|}
\hline
Bone Name & \multicolumn{2}{c|}{Bone Landmarks} & Markers \\ \cline{2-3}
 & Name & Description & \\ \hline
Sacrum & SaCT & Sacral Coccyx Tip & SACR\\ \cline{2-3}
& SaMC & Sacral Median Crest &\\ \cline{2-3}
& SaRSN & Sacral Right Superior Sacral Notch &\\ \hline
Pelvis R & RPASIS & Right Pelvis Anterior Superior Iliac Spine & RASI\\ \cline{2-3}
& RPPSIS & Right Pelvis Posterior Superior Iliac Spine &\\ \cline{2-3}
& RPIT & Right Pelvis Ischial Tuberosity &\\ \hline
Femur R & RFmH & Right Femur Head & RTHI\\ \cline{2-4}
& RFmGT & Right Femur Greater Trochanter & RKNE\\ \cline{2-4}
& RFmMC & Right Femur Medial Condyle & RKNE\_md\\ \cline{2-3}
& RFmLNP & Right Femur Lateral Near to Patella &\\ \cline{2-3}
& RFmLC & Right Femur Lateral Condyle &\\ \hline
Patella R & RPaAP & Right Patella Apex &\\ \hline
Tibia R & RTMC & Right Tibial Medial Condyle & RTIB\\ \cline{2-4}
& RTLT & Right Tibial Tuberosity & RANK\\ \cline{2-4}
& RTMM & Right Tibial Medial Malleolus & RANK\_md\\ \cline{2-3}
& RTLBE & Right Tibial Lateral Bottom Epiphysis &\\ \hline
Fibula R & RFbHM & Right Fibula Head Medial Point &\\ \cline{2-3}
& RFbH & Right Fibula Head &-\\ \cline{2-3}
& RFbLM & Right Fibula Lateral Malleolus &\\ \cline{2-3}
& RFbMM & Right Fibula Medial Malleolus &\\ \hline
Talus R & RFtTIP & Right Foot Talus Inferior Posterior &-\\ \hline
Calcaneus R & RFtCMP  & Right Foot Calcaneous Medial Posterior & RANK, RANK\_md\\ \cline{2-4}
& RFtCLP & Right Foot Calcaneus Lateral Posterior & RTOE, RHEEL\\ \hline
Navicular R & - & - &-\\ \hline
Cuboid R & - & - &-\\ \hline
Medial Cuneiform R & - & - & -\\ \hline
Intermediate Cuneiform R & - & - &-\\ \hline
Lateral Cuneiform R & - & - &-\\ \hline
$1^{st}$ Metatarsal R & RFt1MA & Right Foot 1st Metatarsal Anterior &-\\ \hline
$2^{nd}$-$4^{th}$ Metatarsal R & - & - &-\\ \hline
$5^{th}$ Metatarsal R & RFt5MA & Right Foot 5th Metatarsal Anterior &-\\ \hline
$1^{st}$-$5^{th}$ Proximal Phalanx R & - & - &-\\ \hline
$2^{nd}$-$5^{th}$ Middle Phalanx R & - & - &-\\ \hline
$1^{st}$-$5^{th}$ Distal Phalanx R & - & - &-\\ \hline
\end{tabular}
\end{table}

\end{document}